\documentstyle[twoside,fleqn,espcrc2,psfig]{article}
\newcommand{\lsim}{\raisebox{0.3mm}{\em $\, <$} 
\hspace{-3.3mm} \raisebox{-1.8mm}{\em $\sim \,$}}

\def\lsim{\raise0.3ex\hbox{$\;<$\kern-0.75em\raise-1.1ex
\hbox{$\sim\;$}}}
\def\gsim{\raise0.3ex\hbox{$\;>$\kern-0.75em\raise-1.1ex
\hbox{$\sim\;$}}}

\begin{document}
\topmargin = -1.0cm

\title{
Measuring CP violation by low-energy medium-baseline neutrino 
   oscillation experiments
\thanks{Talk presented by H. Nunokawa based on the work in 
Ref.[1] at ``NuFact'00'' workshop, 
Monterey, CA, USA, May 22-26, 2000.}}

\author{Hisakazu Minakata$^{1,2}$ and Hiroshi Nunokawa$^3$ \\
        {\ }\\
$^1$Department of Physics, Tokyo Metropolitan University \\
1-1 Minami-Osawa, Hachioji, Tokyo 192-0397, Japan, and \\
\vglue 0.1cm
$^2$Research Center for Cosmic Neutrinos, 
Institute for Cosmic Ray Research, \\ 
University of Tokyo, Kashiwa, Chiba 277-8582, Japan \\
\vglue 0.1cm
$^3$Instituto de F\' {\i}sica Gleb Wataghin,
Universidade Estadual de Campinas, UNICAMP\\    
13083-970 -- Campinas, Brazil.}

\begin{abstract}
In this talk we discuss the possibility to measure CP violation 
in neutrino oscillation experiments using the 
neutrino beam with energy which is lower ($E_\nu \gsim 100$ MeV) 
than the one usually considered (typically $> 1$ GeV) 
in accelerator experiments. 
The advantage of using such lower energy neutrino beam is 
that despite the smaller detection cross sections, 
the effect of CP violation is larger and the optimal length 
of baseline can be rather short, 30-50 km, being free from matter 
effect contamination. 
\end{abstract}

\maketitle


\newpage
Very impressive results in atmospheric neutrino observation 
from the SuperKamiokande experiment \cite {SKatm} and 
the persistent discrepancy between the observed and 
the calculated flux of solar neutrinos \cite {solar} 
provide the strongest indications for neutrino oscillations 
induced by neutrino masses and lepton flavor mixing. 
Determination of all the mixing parameters, in particular the  
CP violating Kobayashi-Maskawa phase \cite{KM}, is one of 
the most challenging goals in particle physics. 

In this talk, we consider the possibility of measuring  
the effect of leptonic CP violation (or equivalently T violation) 
in neutrino oscillation~\cite{early,KPTV,MN97,MN98,cpv} experiments 
within the standard three neutrino flavor framework. 
We will end up with the proposal of medium baseline experiments 
with much lower energy ($E_\nu \gsim 100$ MeV) 
neutrino beam than the ones usually considered 
(typically $E_\nu > 1$ GeV) in accelerator experiments. 

Let us consider the neutrino mass squared differences 
suggested by the atmospheric neutrino observation \cite{SKatm}
and the MSW solutions \cite {MSW} to the solar 
neutrino problem. 
In the standard three-flavor neutrino mixing scheme, 
all the independent mass squared differences are 
exhausted if we assume the oscillation interpretations 
of the atmospheric as well as solar neutrino data. 
They are, from the atmospheric neutrinos
$\Delta m_{13}^2 \approx \Delta m_{23}^2 = 
\Delta m_\textrm{atm}^2 \simeq (2-5)\times 10^{-3}\ \textrm{eV}^2$ 
and from the solar neutrino data, 
$\Delta m_{12}^2 = \Delta m_\textrm{solar}^2 \simeq 
(4 - 10)\times10^{-6}\ \textrm{eV}^2\ \textrm{(SMA)},\ 
(2 - 20)\times 10^{-5}\ \textrm{eV}^2\ \textrm{(LMA)}$ or 
$(6 - 20)\times 10^{-8}\ \textrm{eV}^2\ \textrm{(LOW)}$
where SMA, LMA, and LOW denote the small mixing angle, 
large mixing angle and the low $\Delta m^2$ MSW solutions, 
respectively \cite{MSWfit,FLMP}. 
We use the notation $\Delta m^2_{ij} \equiv m_j^2 - m_i^2$
in this report. 

Many authors including us examined the question of how 
to separate the genuine CP violating effect due to the leptonic 
Kobayashi-Maskawa phase from the fake one induced by matter effect 
\cite{MN97,MN98,cpv}. 
Here we take a simple alternative strategy to look for 
a region of parameters in which the matter effect is 
"ignorable" in the first approximation. 

Let us define the flavor mixing matrix $U$ as 
$\nu_{\alpha}=U_{\alpha i}\nu_i$, where 
$\nu_{\alpha}(\alpha=e,\mu, \tau)$ and $\nu_i(i=1,2,3)$
stand for the gauge and the mass eigenstates, respectively.
We take for convenience the representation of $U$ as 
\begin{equation}
U = e^{i\lambda_7 \theta_{23}} \Gamma_{\delta}e^{i \lambda_5\theta_{13}}
e^{i\lambda_2\theta_{12}} 
\end{equation}
where $\lambda_i$ are SU(3) Gell-Mann's matrix and 
$\Gamma_{\delta} = diag (1, 1, e^{i\delta})$ with 
$\delta$ being the CP violating phase. 

Let us first neglect the possible matter effect 
and consider the CP violating effect in vacuum.  
Under the mass difference hierarchy $\Delta m_{13}^2 \gg \Delta m_{12}^2$ 
which is implied by the solar and atmospheric neutrino data, 
the neutrino oscillation probability in vacuum can be written as
\begin{eqnarray}
P(\nu_\beta \rightarrow \nu_\alpha) 
&=& 4 |U_{\alpha 3}|^2|U_{\beta 3}|^2 
\sin^2 \left(\frac{\Delta_{13}L}{2}\right)\nonumber\\ 
&&  \hskip -2.cm =
 -4 \mbox{Re} [U_{\alpha 1}U_{\alpha 2}^*U_{\beta 1}^*U_{\beta 2}]
\sin^2 \left(\frac{\Delta_{12}L}{2}\right)\nonumber\\ 
&& \hskip -2cm  - 2J \sin (\Delta_{12}L) 
[1 - \cos(\Delta_{13}L)] \nonumber\\ 
&& \hskip -2.cm  + 4J \sin (\Delta_{13}L) \sin^2 (\Delta_{12}L), 
\label{probvac}
\end{eqnarray}
where 
$\Delta_{ij} \equiv  \frac{\Delta m^2_{ij}}{2E} $
and $L$ is the distance traveled by neutrinos. 
Here the effect of CP violation comes in through 
$J$ in eq. (\ref{probvac}) which is defined as 
\begin{equation}
J_{\alpha\beta; i,j} \equiv \mbox{Im}
[U_{\alpha i}U_{\alpha j}^*U_{\beta i}^*U_{\beta j}]
\end{equation}
as it is the unique 
(in three-flavor mixing scheme) measure for CP violation 
as first observed by Jarlskog \cite{Jarlskog} in the case of quark mixing. 
With our parametrization of mixing matrix, it takes the form 
$J = \pm c_{12}s_{12}c_{23}s_{23}c_{13}^2s_{13} \sin{\delta}$, 
where the sign is positive for $(e, \mu)$ and (1, 2) and $+(-)$ 
corresponds to their (anti-) cyclic permutations of $(\alpha, \beta)$ 
and $(i, j)$.

We first note that from the expression of $J$, if any one of 
the mixing angles is extremely small or very close to $\pi/2$ 
there is little hope in detecting the CP violation
even if $\delta$ is large. 
For this reason, we will not deal with the case of SMA MSW 
solar neutrino solution. 

We also notice that the effect of CP violation is larger 
if the quantity 
\begin{eqnarray}
\Delta_{12} L &=& 0.026
\left(\frac{\Delta m_{12}^2}{10^{-5}\ \mathrm{eV}^2}\right)
\left(\frac{L}{100\ \mathrm{km}}\right) \nonumber \\
&& \left(\frac{E}{100\ \mathrm{MeV}}\right)^{-1}, 
\label{deltaL}
\end{eqnarray}
is larger. 
{}From this we see that larger values of 
$\Delta m_{12}^2$, lower $E_\nu$ and larger $L$ is preferrable. 
For this reason we consider largest plausible value 
of $\Delta m_{12}^2$ coming from LMA MSW solution 
(which is chosen among the suggested solutions) 
to the solar neutrino problem and lowest possibly 
practical energy $\sim$ 100 MeV which can be provided 
by accelerator experiments. 
We note that larger distance $L$ 
is preferrable just in terms of the probability but 
it has to be determined such that the observables of 
the CP violating effect which is proportional to the 
expected number of events (not the probability itself) 
is significant. (See our results below).  

Next let us consider the matter effect.
We first note that at neutrino energy of $\sim$ 100 MeV there is a 
hierarchy among the relevant energy scales;
\begin{eqnarray}
&&\Delta_{13} \equiv \frac{\Delta m_{13}^2}{2E} \sim 10^{-11} 
\ \mathrm{eV} \nonumber\\
&&\gg a \sim
\frac{\Delta m_{12}^2}{2E} \equiv \Delta_{12} \sim 10^{-13} 
\ \mathrm{eV}, 
\label{hierarchy}
\end{eqnarray}
where we assume 
$\rho = 2.7 \mathrm{g/cm}^3$ and 
$\Delta m^2_{12} \sim (3.0-4.0) \times 10^{-5}$ eV$^2$ 
$\sin^2 2\theta_{12} \sim 0.8$ 
assuming LMA MSW solution~\cite{MSWfit} and 
$\Delta m^2_{13} \sim \times 10^{-3}$ eV$^2$ 
$\sin^2 2\theta_{23} \sim 1$ assuming $\nu_\mu-\nu_\tau$ 
atmospheric neutrino oscillation solution. 

Thanks to the mass hierarchy (\ref{hierarchy}) we can formulate 
the perturbation theory and 
it is straightforward to compute the 
neutrino oscillation probabilities 
$P(\nu_{\mu} \rightarrow \nu_e)$ as well as 
$P(\bar{\nu}_{\mu} \rightarrow \bar{\nu}_e)$ 
under the adiabatic approximation~\cite{KS99,Yasuda99,MN00}.  
For example, the appearance probability 
$P(\nu_{\mu} \rightarrow \nu_e)$ 
reads 
\begin{eqnarray}
P(\nu_{\mu} \rightarrow \nu_e) 
&=& 
4 s_{23}^2 c_{13}^2 s_{13}^2 \sin^2 (\frac{1}{2}\Delta_{13}L) \nonumber\\
&& 
\hskip -2cm + c_{13}^2\sin 2\theta_{12}^M \left[ 
(c_{23}^2 - s_{23}^2 s_{13}^2) \sin 2\theta_{12}^M 
\right. \nonumber\\
&& 
\hskip -2cm 
\left.
+ 2c_{23}s_{23}s_{13}\cos{\delta}\cos{2\theta_{12}^M} \right]\nonumber\\
&& 
\hskip -2.5 cm \times \sin^2
\left[\frac{1}{2} \xi \Delta_{12}L \right]
 - 2J_M (\theta_{12}^M, \delta)\sin \left[ \xi\Delta_{12}L \right],
\label{probnu}.
\end{eqnarray}
with 
\begin{equation}
\sin 2\theta_{12}^M \equiv \sin 2\theta_{12}/\xi 
\end{equation}
where
\begin{equation}
\xi \equiv 
\sqrt{(\cos 2\theta_{12} -
\frac{a}{\Delta_{12}}c_{13}^2)^2 + \sin^2 2\theta_{12}}
\end{equation}
and $J_M$ is the matter enhanced Jarlskog factor, 
$J_M (\theta_{12}^M, \delta) 
=\cos \theta_{12}^M \sin \theta_{12}^M c_{23}s_{23}c_{13}^2
s_{13}\sin \delta$, and we have averaged the rapidly oscillating 
piece driven by $\Delta_{13}$ in the CP violating term.
The antineutrino transition probability 
$P(\bar{\nu}_{\mu} \rightarrow \bar{\nu}_e)$
is given by the same expressions as above but replacing 
$a$ and $\delta$ by $-a$ and $- \delta$, respectively. 

Let us note that if $L <$ 1,000 km or so, 
the approximation $\sin (\xi\Delta_{12}L) \simeq \xi\Delta_{12}L$ 
is valid and 
the expressions of the oscillation probabilities approximately 
reduce to those in the vacuum because 
\begin{equation}
\sin 2\theta_{12}^M (\mbox{or}, \bar{\theta}_{12}^M)
\xi \Delta_{12} = \sin 2\theta_{12} \Delta_{12}. 
\end{equation}

\begin{figure}[ht]
\vglue 0.3cm 
\centerline{\protect\hbox{
\psfig{file=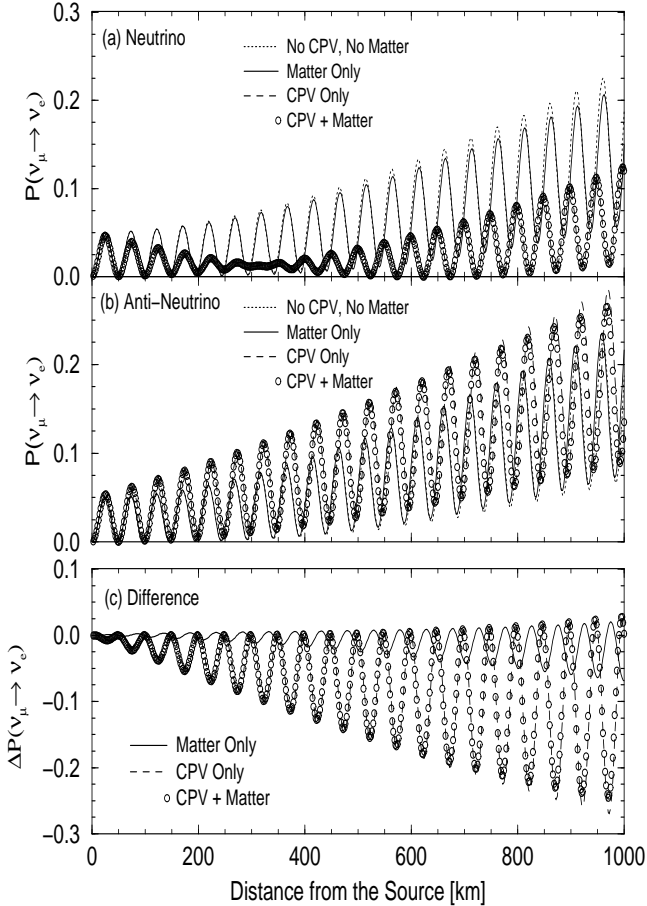,height=11cm,width=9.5cm}
}}
\vglue -0.5cm 
\caption{
Oscillation probability for (a) neutrinos, $P(\nu_\mu\to\nu_e)$,  
(b) anti-neutrinos, $P(\bar{\nu}_\mu\to\bar{\nu}_e)$,  
and (c) their difference,
$\Delta P(\nu_\mu\to\nu_e) \equiv 
P(\nu_\mu \to \nu_e) - P(\bar{\nu}_\mu\to\bar{\nu}_e)$ 
with fixed neutrino energy $E_\nu = $ 60 MeV, 
are plotted as a function of distance from the source. 
The mixing parameters are fixed to be 
$\Delta m^2_{13} = 3 \times 10^{-3}$ eV$^2$, 
$\sin^22\theta_{23} = 1.0$, 
$\Delta m^2_{12} = 2.7 \times 10^{-5}$ eV$^2$, 
$\sin^22\theta_{12} = 0.79$,
$\sin^22\theta_{13} = 0.1$ and 
$\delta = \pi/2$. 
We take the matter density as $\rho = 2.72$ g/cm$^3$ and 
the electron fraction as $Y_e$ = 0.5. 
}
\label{Fig1}
\vglue -0.5cm 
\end{figure}

Using the mixing parameters described in the caption of Fig. 1,
we verify by our numerical computations 
that the matter effect in fact cancels out.
In Fig. \ref{Fig1} we plot the oscillation probabilities for neutrino 
and anti-neutrino, and their difference, 
$\Delta P \equiv P(\nu_{\mu} \rightarrow \nu_e) - 
P(\bar{\nu_{\mu}} \rightarrow \bar{\nu_e})$, 
as a function of 
distance from the source with the neutrino energy $E = 60$ MeV, 
for both 
$\nu_{\mu} \rightarrow \nu_e$ and $\bar{\nu_{\mu}} \rightarrow \bar{\nu_e}$,
where it corresponds to the resonance energy in the neutrino channel.
Although we sit on at the resonance energy of neutrino channel 
the features in the resonant neutrino flavor conversion cannot be 
traced in Fig. \ref{Fig1}, but rather we observe the one very 
much similar to the vacuum oscillation even for $L$ as large
as 1000 km. 
%

Now we discuss the possible experiments which utilize 
the imitating vacuum mechanism to measure leptonic CP violation.
Measurement of CP violation at a few $\%$ (or smaller) 
level at neutrino oscillation experiments at $E \simeq 100$ MeV 
leaves practically the unique channel $\nu_{\mu} \rightarrow \nu_e$. 
Experimentally, what is relevant is the the number 
of events not the oscillation probability itself. 
Here, as a measure of CP violation, 
we consider the ratio of the expected number
of events due to $\nu_{\mu} \rightarrow \nu_e$ 
and $\bar{\nu_{\mu}} \rightarrow \bar{\nu_e}$ 
reactions.

In such low energy appearance experiment 
we have the following difficulties; (1) smaller cross sections 
(2) lower flux due to larger beam opening 
angle, $\Delta \theta \simeq 1$ (E/100 MeV)$^{-1}$ radian. 
Therefore, we are invited to the idea of the baseline as short as 
possible, because the luminosity decreases as $L^{-2}$ as baseline 
length grows. 
However, from Fig. 1 we see that longer the 
baseline, the larger the CP violating effect 
and it is expected that there exist some optimal 
distance which gives the signal most (statistically) significant.

Detection of low-energy neutrinos at better than a few $\%$ 
level accuracy requires supermassive detectors. Probably the best 
thinkable detection apparatus is the water Cherenkov detector of 
SuperKamiokande type. 
Let us estimate the expected number of events 
at a megaton detector placed at $L=100$ km. We assume that the neutrino 
beam flux 10 times as intense as (despite the difference in energy) 
that of the design luminosity in K2K experiment \cite {Nishikawa}. 
We note that in the future it is expected that even an 100 times 
more intense proton flux than KEK-ps seems possible at Japan 
Hadron Facility \cite {Mori}.

The dominant $\nu_e$-induced reaction in water at around $E=100$ MeV 
is not the familiar $\nu_e-e$ elastic scattering but the reaction on 
$^{16}O$,  $\nu_e ^{16}O \rightarrow e^{-} F$ \cite {Haxton}.
The cross section of the former reaction is about 
$\sigma(\nu_e e \rightarrow \nu_e e) =
0.93 \times 10^{-42}
\displaystyle\left(\frac{E}{100\ \mathrm{MeV}}\right) 
\mathrm{cm}^2$, while the latter 
is 
$\sigma(\nu_e ^{16}O \rightarrow e^{-} F) \simeq 10^{-39} 
\mathrm{cm}^2$ at $E=100$ MeV \cite {Haxton}, 
which is a factor of 1000 times larger.
However, since the number of oxygen in water is 1/10 of the 
number of electrons, net the number of events due to the reaction 
$\nu_e ^{16} O \rightarrow e^{-} F$ is larger than that of 
$\nu_e e$ elastic scattering by a factor of 100. 
The neutrino flux at the detector located at $L=250$ km is, 
by our assumption, 10 times more intense 
than the neutrino flux at SuperKamiokande in K2K experiment.
The latter is, roughly speaking, 
$3 \times 10^{6} 
\displaystyle\left(\frac{\mathrm{POT}}{10^{20}}\right)$ 
cm$^{-2}$ where 
POT stands for proton on target.
Therefore, the expected number of events $N$ assuming 100 $\%$ 
conversion of $\nu_{\mu}$ to $\nu_e$ is given by 
\begin{equation}
N \simeq 6300
\ \displaystyle\left(\frac{L}{100\ \mathrm{km}}\right)^{-2}
\displaystyle\left(\frac{V}{1\ \mathrm{Mton}}\right) 
\displaystyle\left(\frac{\mathrm{POT}}{10^{21}}\right). 
\label{eventNo}
\end{equation}

In the antineutrino channel, the dominant reaction is 
$\bar{\nu}_e p \rightarrow e^+ n$ with cross section
$\sigma \simeq 0.4 \times 10^{-39} \mathrm{cm}^2$ at $E=100$ MeV. 
The event number due to this reaction, assuming the same 
flux of $\bar{\nu}_{\mu}$ as $\nu_{\mu}$, is similar to that of 
(\ref{eventNo}) because the cross section is about half but 
there are two free protons per one oxygen.  
There is additional oxygen reaction 
$\bar{\nu}_e ^{16}O \rightarrow e^{+} N$ 
with approximately factor 3 smaller cross section than that 
of $\nu_e ^{16}O$ \cite {Haxton}.

\begin{figure}[ht]
\vglue 0.8cm 
\centerline{\protect\hbox{
\psfig{file=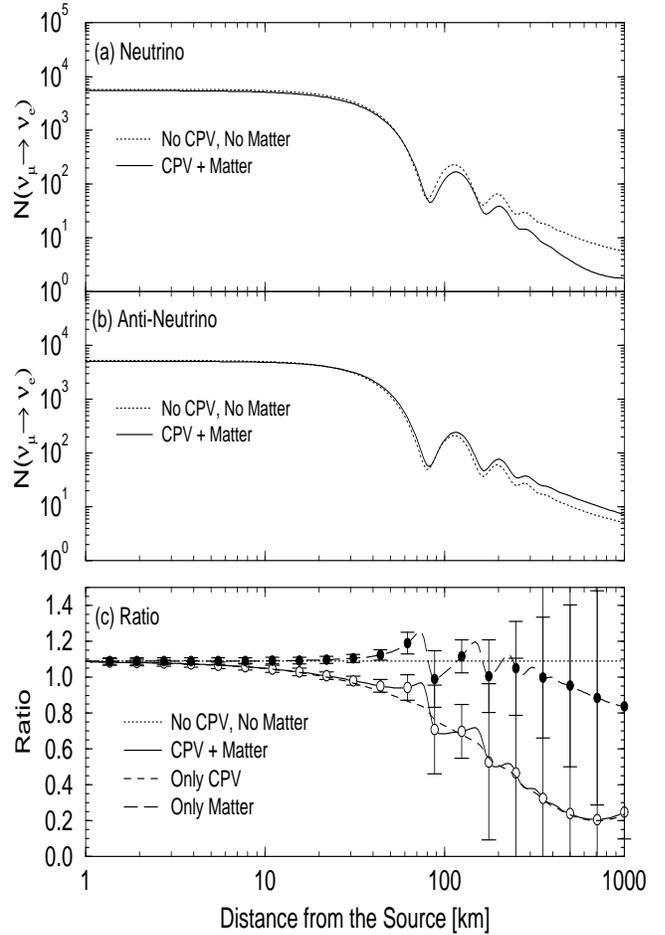,height=11cm,width=9.5cm}
}}
\vglue -0.2cm 
\caption{
Expected number of events for 
(a) neutrinos, $N(\nu_\mu\to\nu_e)$,  
(b) anti-neutrinos, $N(\bar{\nu}_\mu\to\bar{\nu}_e)$,  
and (c) their ratio 
$R\equiv N(\nu_\mu \to \nu_e)/N(\bar{\nu}_\mu\to\bar{\nu}_e)$ 
with a Gaussian type neutrino energy beam with 
$\langle E_\nu \rangle = $ 100 MeV with $\sigma$ = 10 MeV 
are plotted as a function of distance from the source. 
Neutrino fluxes are assumed to vary as $\sim 1/L^2$ 
in all the distance range we consider. 
The mixing parameters as well as the electron number density 
are fixed to be the same as in Fig. 1. 
The error bars are only statistical.
}
\label{Fig2}
\vglue -0.6cm 
\end{figure}

In order to estimate the optimal distance, we compute the expected 
number of events in neutrino and anti-neutrino channels as well as 
their ratios as a function of distance taking into account of 
neutrino beam energy spread. 
For definiteness, we assume that the average energy of neutrino 
beam $\langle E \rangle$ = 100 MeV and beam energy spread of 
Gaussian type with width $\sigma_E = 10$ MeV. 
We present our results in Fig. \ref{Fig2}. 

We see from this plot that taking the error (only statistical 
one for simplicity) into account, $L$ must not be too short 
or too long and there exist on 
optimal distance which is rather short, $L \sim 30-50 $ km, 
much shorter than that of the current (or future) long-baseline 
neutrino oscillation experiments (see Fig. 2).
At such distance the significance of the CP violating
signal is as large as 3-4 $\sigma$.

We conclude that the experiment is quite feasible under such intense 
neutrino beam and a megaton detector. 
Fortunately, the possibility of constructing a megaton water 
Cherenkov detector is already discussed by 
the experimentalists \cite{Nakamura}.

Some final remarks are in order. 

\vglue 0.1cm 
\noindent
1) For the other MSW solutions such as SMA or LOW to the
solar neutrino problem, due to the much smaller mixing
angle and/or $\Delta m^2$, with the same experimental 
conditions, the expected number of events is 
smaller than at least a factor of 100 and therefore, 
the experiment does not appear feasible with the 
same apparatus. 

\vglue 0.1cm 
\noindent
2) An alternative way of measuring CP violation is 
the multiple detector method \cite{MN97} which may be 
inevitable if either one of $\nu$ or $\bar{\nu}$ beam 
is difficult to prepare. 
It utilizes the fact that the first, the second, and the third terms 
in the oscillation probability (\ref{probnu}) have different $L$ 
dependences, $\sim L$-independent 
(after averaging over energy spread of the neutrino beam), 
$\sim L^2$, and $\sim L$, respectively, in the linear approximation.

\vglue 0.1cm 
\noindent
3) Intense neutrino beams from muon storage ring at low 
energies, proposed as PRISM, Phase-Rotation Intense Secondary 
Mesons \cite {Kuno}, 
could be an ideal source for neutrinos for the experiment 
discussed in this report.
Of course, it requires identification of 
$\bar{\nu_e}$ from $\nu_e$ by some methods, e.g., 
by adding Chlorine ($^{35}$Cl) into the Water Cherenkov detector 
to make it sensitive to the characteristic $\sim 8$ MeV $\gamma$ 
rays arising from the absorption of neutron into the Chlorine 
followed by $\bar{\nu}_e p \rightarrow e^+ n$ reaction. 

\vglue 0.1cm 
\noindent
4) We would like to urge experimentalists to think more about 
the better supermassive detection apparatus than water Cherenkov 
for highly efficient and accurate measurement of low energy 
neutrinos.

See Ref.~\cite{MN00} for more detailed discussions on this
work.

\section*{Acknowledgments}

We thank Takaaki Kajita, Masayuki Nakahata and Kenzo Nakamura 
for informative discussions on detection of low energy neutrinos.
This work was supported by the Brazilian funding agency 
Funda\c{c}\~ao de Amparo \`a Pesquisa do Estado de S\~ao Paulo (FAPESP), 
and by the Grant-in-Aid for Scientific Research in Priority Areas No. 
11127213, Japanese Ministry of Education, Science, Sports and Culture.


\end{document}